\documentclass[a4paper,
               bibtex,     
               keeplastbox,   
               ]{jacow}
\usepackage{pdfpages,multirow,ragged2e} %
\makeatletter%
	\ifboolexpr{bool{xetex}}
	 {\renewcommand{\Gin@extensions}{.pdf,%
	                    .png,.jpg,.bmp,.pict,.tif,.psd,.mac,.sga,.tga,.gif,%
	                    .eps,.ps,%
	                    }}{}
\makeatother

\ifboolexpr{bool{xetex} or bool{luatex}} 
 {}                                      
 {\usepackage[utf8]{inputenc}}           

\usepackage[UKenglish]{babel}

\ifboolexpr{bool{jacowbiblatex}}%
 {%
  \addbibresource{references.bib}
 }{}
\newcommand{\diff}{\mathrm{d}}

\DeclareFontFamily{U}{BOONDOX-calo}{\skewchar\font=45 }
\DeclareFontShape{U}{BOONDOX-calo}{m}{n}{
  <-> s*[1.05] BOONDOX-r-calo}{}
\DeclareFontShape{U}{BOONDOX-calo}{b}{n}{
  <-> s*[1.05] BOONDOX-b-calo}{}
\DeclareMathAlphabet{\mathcalboondox}{U}{BOONDOX-calo}{m}{n}
\SetMathAlphabet{\mathcalboondox}{bold}{U}{BOONDOX-calo}{b}{n}
\DeclareMathAlphabet{\mathbcalboondox}{U}{BOONDOX-calo}{b}{n}
\newcommand{\calE}{\mathcalboondox{E}}

\let\OLDthebibliography\thebibliography
\renewcommand\thebibliography[1]{
  \OLDthebibliography{#1}
  \setlength{\parskip}{0.0pt}
  \setlength{\itemsep}{0pt plus 0.3ex}
}

\usepackage[scaled]{DejaVuSansMono}
\usepackage[T1]{fontenc}

\begin{document}

\title{$\gamma\gamma$ Beam-beam Parameter Study for a $\SI{3}{\tera\electronvolt}$ PWFA Linear Collider}

\author{J. B. B. Chen\textsuperscript{1}, D. Schulte, CERN, Geneva, Switzerland \\
		E. Adli, University of Oslo, Oslo, Norway \\
		\textsuperscript{1}also at University of Oslo, Oslo, Norway}
	
\maketitle

\begin{abstract}
    Our beam-beam parameter study using beam-beam simulations and PWFA (particle-driven plasma acceleration) beam parameters indicates that at \SI{3}{\tera\electronvolt}, for examined electron beam lengths ${\SI{2}{\micro\meter}\leq\sigma_z\leq\SI{10}{\micro\meter}}$, the total luminosity, as well as the sharpness of the luminosity spectrum for a $\gamma\gamma$ collider are independent of the beam length of the electron beams used to scatter the photons, given that the hourglass effect is avoided. The total luminosity can consequently be maximised by minimising the horizontal and vertical beta functions $\beta_{x,y}^*$ at the interaction point.
    
    Furthermore, we performed background studies in GUINEA-PIG where we considered the smallest currently achievable $\beta_{x,y}^*$ combined with PWFA beam parameters. Simulations results show that our proposed parameter set for a \SI{3}{\tera\electronvolt} PWFA $\gamma\gamma$ collider is able to deliver a total luminosity significantly higher than a $\gamma\gamma$ collider based on CLIC parameters, but gives rise to more background particles.
\end{abstract}

\section{Introduction}
In the blow-out regime of PWFA (particle-driven plasma wakefield acceleration), a dense ultra-relativistic drive beam is used to excite a plasma wake, where plasma electrons are expelled from the region close to the propagation axis, leaving only positively charged ions behind to form a plasma ion bubble cavity. Inside the plasma ion cavity accelerating gradients in the multi-\SI{}{\giga\volt/\meter} level \cite{Blumenfeld} can be used to accelerate a trailing main beam consisting of electrons or positrons.

However, due to the challenges of accelerating positrons using PWFA, a $\gamma\gamma$ collider, where laser photons are scattered off ultra-relativistic electrons accelerated by PWFA, has been proposed as an alternative to a PWFA e\textsuperscript{$+$}e\textsuperscript{$-$} linear collider. A $\gamma\gamma$ collider has the potential to provide a higher luminosity than a  e\textsuperscript{$+$}e\textsuperscript{$-$} collider due to the absence of beamstrahlung in $\gamma\gamma$ collisions, while also offering an attractive discovery potential \cite{Telnov_gg_Physics_Goal}.

A previous parameter study on a \SI{1.5}{\tera\electronvolt} PWFA accelerator \cite{Chen2020modeling} derived a parameter set that can provide reasonable stability, energy spread and efficiency for electron acceleration. This beam-beam study used the electron main beam parameter set in \cite{Chen2020modeling} as a basis to optimise beam parameters in the interaction point (IP) for a TeV-scale $\gamma\gamma$ collider.

\section{$\gamma\gamma$ collider principles}

The principles of a $\gamma\gamma$ collider are outlined in \cite{Telnov_gammagammaPrinciples, Ginzburg_gammagamma}. We will here give a brief summary of the basic concepts.

Inverse Compton scattering of laser photons on relativistic electrons is considered the most efficient method to produce the required high-energy photons. A short distance before the interaction point, a high-energy electron beam collides with a laser beam in the conversion region. After the Compton scattering process, the back-scattered photons have acquired a large fraction of the incident electrons' energy, and follow the direction of the incident electrons to be collided in the interaction point with other back-scattered photons. 


\subsection{Inverse Compton Scattering Kinematics}
\begin{figure}[!htb]
    \centering
    \resizebox{0.6\columnwidth}{!}{
            \def\svgwidth{0.3\textwidth}
            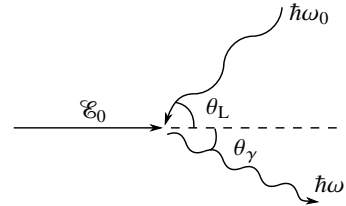
        }
    \caption{Inverse Compton scattering. A photon with energy $\hbar\omega_0$ scatters off an ultra-relativistic electron with energy $\calE_0$. The photon's energy after scattering is $\hbar\omega$.}
    \label{fig:inverseCompton}
\end{figure}

In the conversion region, laser photons are scattered off ultra-relativistic electrons at small collision angles $\theta_\mathrm{L}$ as shown in figure \ref{fig:inverseCompton}. Before scattering, the electron has energy $\calE_0$ and momentum $p_0\approx\calE_0/c$, while that of the photon is $\hbar\omega_0$. The inner product of the total four-momentum is then given by
\begin{equation}
    P^\mu P_\mu \approx m_\mathrm{e}^2c^2 + \frac{4\calE_0\hbar\omega_0}{c^2}\cos^2\frac{\theta_\mathrm{L}}{2}.
\end{equation}

We define the dimensionless invariant energy parameter \begin{equation}
    x = \frac{4\calE_0\hbar\omega_0}{(m_\mathrm{e}c^2)^2}\cos^2\frac{\theta_\mathrm{L}}{2}
\end{equation}
that can be used as a measure for the energy of the electron-photon system. $x$ is related to the centre of mass energy $\sqrt{s}$ by ${s = m_\mathrm{e}^2c^4(1+x)}$.

After collision, the majority of the back-scattered photons move in directions given by small angles $\theta_\gamma$ relative to the trajectory of the incoming electron. The frequency of the back-scattered photon is given by
\begin{equation}
    \omega = \frac{\omega_\mathrm{m}}{1+(\theta_\gamma/\vartheta_0)^2},
\end{equation}
where ${\vartheta_0 = \frac{m_\mathrm{e}c^2}{\calE_0}\sqrt{x+1}}$ and ${\omega_\mathrm{m} = \frac{x}{x+1}\frac{\calE_0}{\hbar}}$ is the maximum photon frequency, which corresponds to the photon being scattered in the same direction as the incoming electron.

\subsection*{Photon Energy Spectrum}
Let ${y=\hbar\omega/\calE_0}$. The energy spectrum of the back-scattered photons is determined by the quantity $(1/\sigma_\mathrm{c})(\diff\sigma_\mathrm{c}/\diff y)$, where $\sigma_\mathrm{c}$ is the total cross section of the Inverse scattering process. The differential cross section for the inverse Compton scattering process is given by \cite{Telnov_gammagammaPrinciples}
\begin{align}
\begin{split}
    \frac{\diff\sigma_\mathrm{c}}{\diff y} &= \frac{2\pi r_\mathrm{e}^2}{x}\left[ \frac{1}{1-y}+1-y - \frac{4y}{x(1-y)}\left(1-\frac{y}{x(1-y)}\right)\right.\\
    &\left.+ \frac{2h_\mathrm{e}\Omega y}{1-y}\left(1-\frac{2y}{x(1-y)}\right)(2-y) \right],
\end{split}
\end{align}
where $r_\mathrm{e}$ is the classical electron radius, $h_\mathrm{e}$ is the helicity of the electron and $\Omega$ is the polarisation of the initial photon. The sharpness of the photon energy spectrum strongly depends on the value of $2h_\mathrm{e}\Omega$, which should ideally be $-1$. In practice, $2h_\mathrm{e}\Omega\approx -0.8$ is a more realistic value, which we adopt in this study.


The energy of the back-scattered photons increases with increasing laser photon energy. However, the  back-scattered photons and laser photons may scatter, and if their energy exceed the threshold for $\mathrm{e}^+\mathrm{e}^-$ pair creation, high-energy photons will be lost, and unwanted $\mathrm{e}^+\mathrm{e}^-$ pairs will be created. The energy of the laser photons is thus restricted by $\mathrm{e}^+\mathrm{e}^-$ pair creation. Four-momentum considerations for $\mathrm{e}^+\mathrm{e}^-$ pair creation gives the upper limits ${x<2+2\sqrt{2}\approx 4.8}$ and ${y_\mathrm{m}\approx 0.83}$. In this study, we set $x=4.8$.



\subsection*{Conversion Efficiency}
The conversion efficiency $\eta_\gamma$, which gives the average number of high-energy (back-scattered) photons $N_\gamma$ per electron, is related to the laser pulse energy $A$ by \cite{Ginzburg_gammagamma}
\begin{equation}
    \eta_\gamma = \frac{N_\gamma}{N_\mathrm{e}} = 1-e^{-A/A_0},
\end{equation}
where $N_\mathrm{e}$ is the total number of electrons in the incoming beam and $A_0=\hbar cl_\mathrm{L}/(2\sigma_\mathrm{c})$ is a parameter depending on the laser beam length $l_\mathrm{L}$.

Large laser pulse energies results in a higher number of high-energy photons, but since electrons have an increased chance of multiple scatterings, this will also lead to more low-energy photons, which will enhance the low-energy part of the spectrum. Thus, there will be a trade-off between the total luminosity and the degree of monochromaticity of the luminosity spectrum. 

\subsection*{Distance between Conversion Region and Interaction Point}
The angle of a back-scattered photon with respect to the direction of the initial electron is given by ${\theta_\gamma = \vartheta_0\sqrt{y_\mathrm{m}/y-1}}$, which decreases with increasing photon energy.

Thus, the contribution to the luminosity from low-energy photons decreases faster over distance compared to high-energy photons. The luminosity spectrum will depend on the distance $d$ between the conversion region and the interaction point. For large distances, the luminosity spectrum will be sharply peaked at high energies, but the total luminosity will decrease. Hence, this represent another trade-off between the total luminosity and the sharpness of the luminosity spectrum.

The dimensionless parameter ${\varrho = d/(\gamma\sigma_y^*)}$ is often used to describe the distance between the conversion region and the interaction point. 

\section{Parameter Study Results}

    
In a previous parameter study \cite{Chen2020modeling} on a $\SI{1.5}{\tera\electronvolt}$ PWFA linear accelerator, we found a parameter set for the main beam with acceptable stability, energy spread and efficiency. This parameter set involves an electron beam with $N=\SI{5e9}{}$ electrons and a rms beam length of $\sigma_z=\SI{5}{\micro\meter}$, which we adopt in this study.

\subsection*{Luminosity Spectrum}

\begin{figure}[!htb]
    \centering
    \includegraphics[width=0.75\columnwidth]{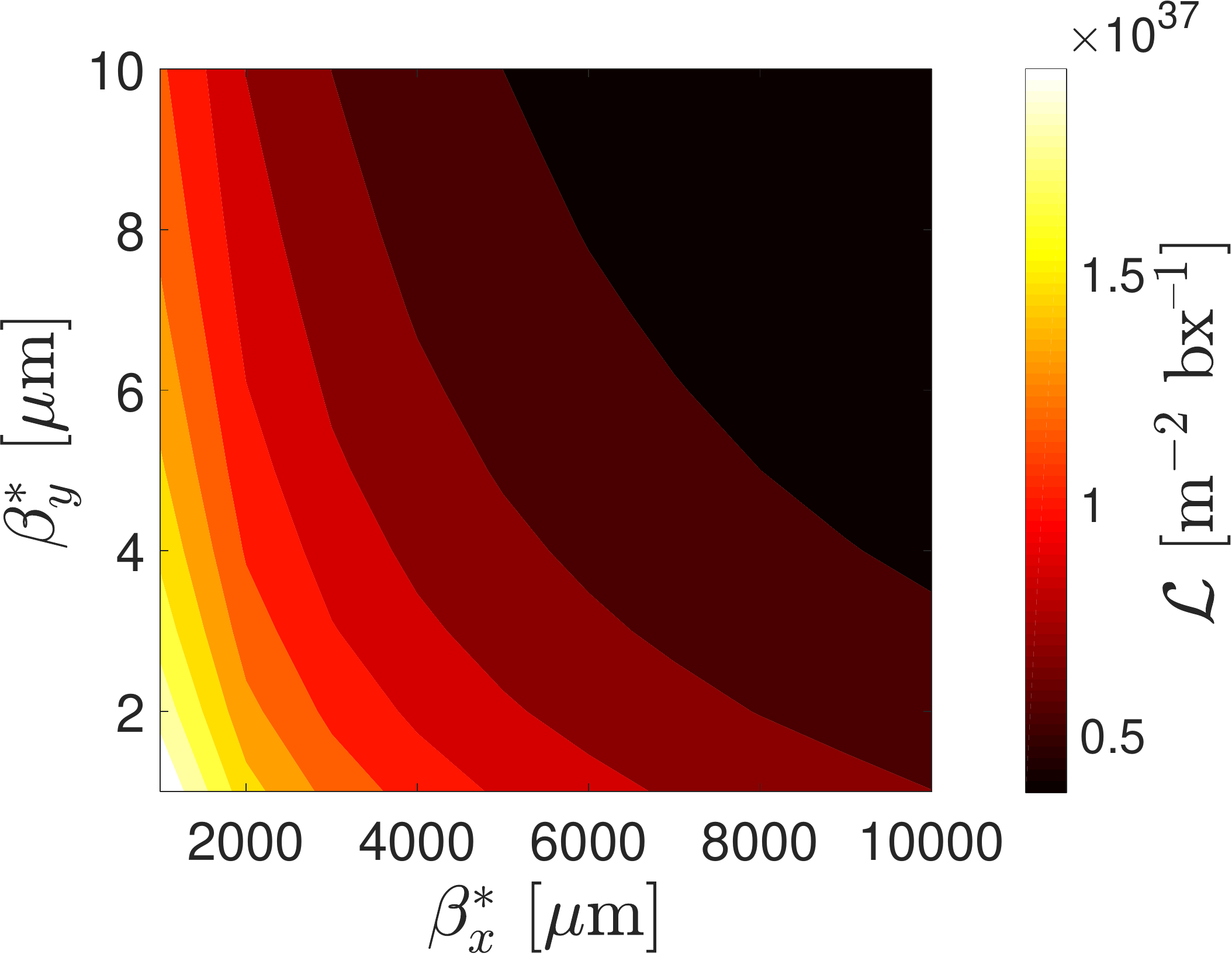}
    \caption{Contour plot of total luminosity $\mathcal{L}$ vs. horizontal and vertical beta function $\beta_x^*$ and $\beta_y^*$ at the interaction point.}
    \label{2020_N5e9_sigmaZ5um_lumigg_contourPlot}
\end{figure}

\begin{figure}[!htb]
    \centering
    \includegraphics[width=0.75\columnwidth]{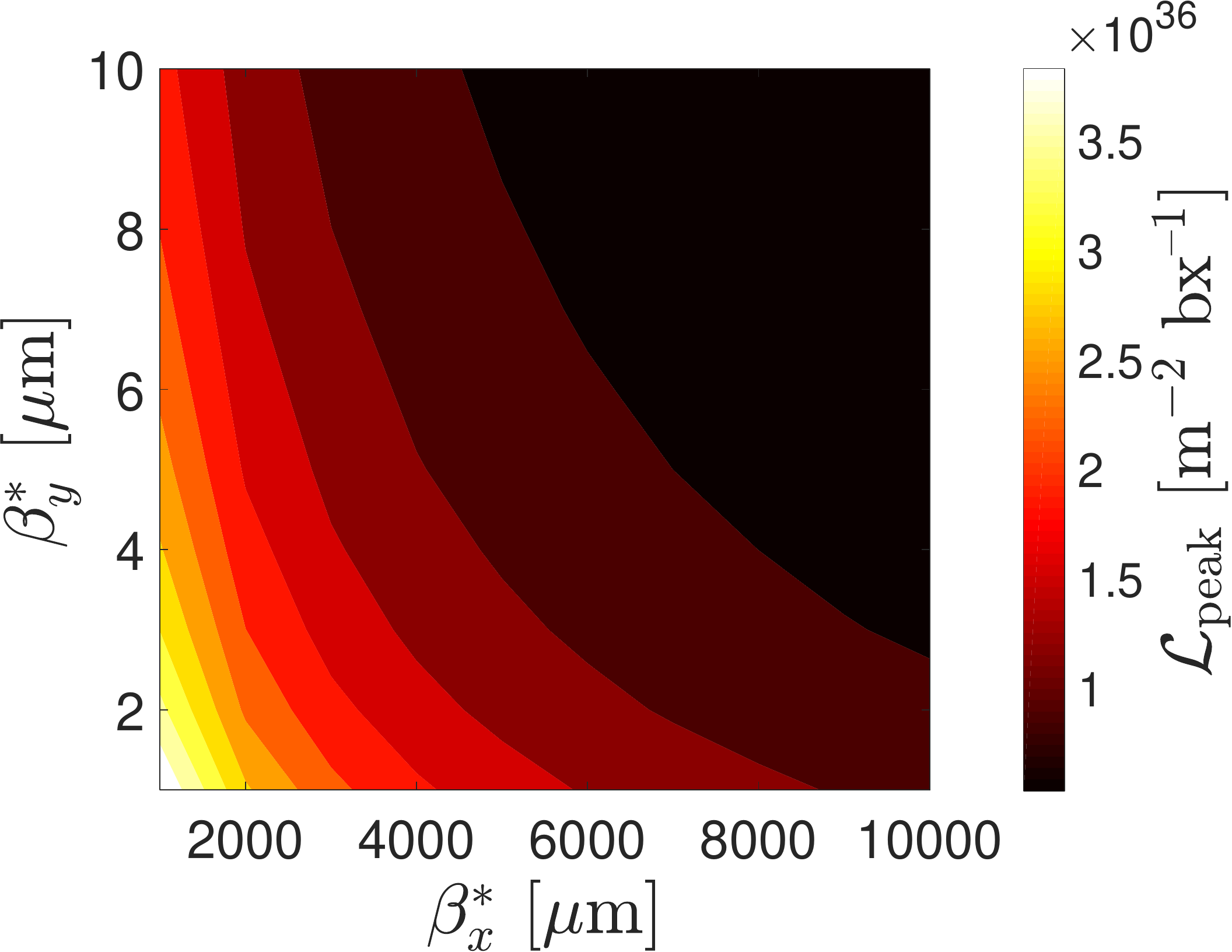}
    \caption{Contour plot of total peak luminosity $\mathcal{L}_\mathrm{peak}$ vs. horizontal and vertical beta function $\beta_x^*$ and $\beta_y^*$ at the interaction point.}
    \label{2020_N5e9_sigmaZ5um_lumiPeakgg_contourPlot}
\end{figure}

Plots of $\mathcal{L}$ and the peak luminosity $\mathcal{L}_\mathrm{peak}$ per beam crossing (bx), defined here as the part of the luminosity corresponding to centre of mass energy $\sqrt{s}>0.80y_\mathrm{m}\sqrt{s_0}$, where $\sqrt{s_0}$ is the nominal centre of mass collision energy, are shown in fig. \ref{2020_N5e9_sigmaZ5um_lumigg_contourPlot} and \ref{2020_N5e9_sigmaZ5um_lumiPeakgg_contourPlot} for $N=\SI{5e9}{}$, $\sigma_z=\SI{5}{\micro\meter}$.

By performing a parameter scan for a \SI{3}{\tera\electronvolt} $\gamma\gamma$ collider in GUINEA-PIG \cite{Schulte_thesis} over a range of horizontal and vertical beta functions $\beta_x^*$, $\beta_y^*$ at the interaction point (IP) and rms beam length $\sigma_z$, we found that the total luminosity $\mathcal{L}$ and the peak luminosity are independent of $\sigma_z$ for short beams in the examined interval ${\SI{2}{\micro\meter}\leq\sigma_z\leq\SI{10}{\micro\meter}}$ \footnote{The hourglass effect was avoided since $\beta_y^*\geq\sigma_z$.}. For e\textsuperscript{$+$}e\textsuperscript{$-$} collisions, a $\sigma_z$ dependence was introduced to the total luminosity through beamstrahlung limitation requirements \cite{Chen2020ee}. This requirement does not exist for $\gamma\gamma$ collisions, so the independence of $\sigma_z$ is expected. Furthermore, both the luminosity and the sharpness of the spectrum are increased with decreasing $\beta_x^*$ and $\beta_y^*$ \footnote{At high energy and larger beam charges, conversion of high energy photons into coherent pairs in the fields of the opposing electron beam can however restrict the horizontal spot size \cite{Telnov_gg_Physics_Goal}.}. 

\begin{figure}[!htb]
    \centering
    \includegraphics[width=0.65\columnwidth]{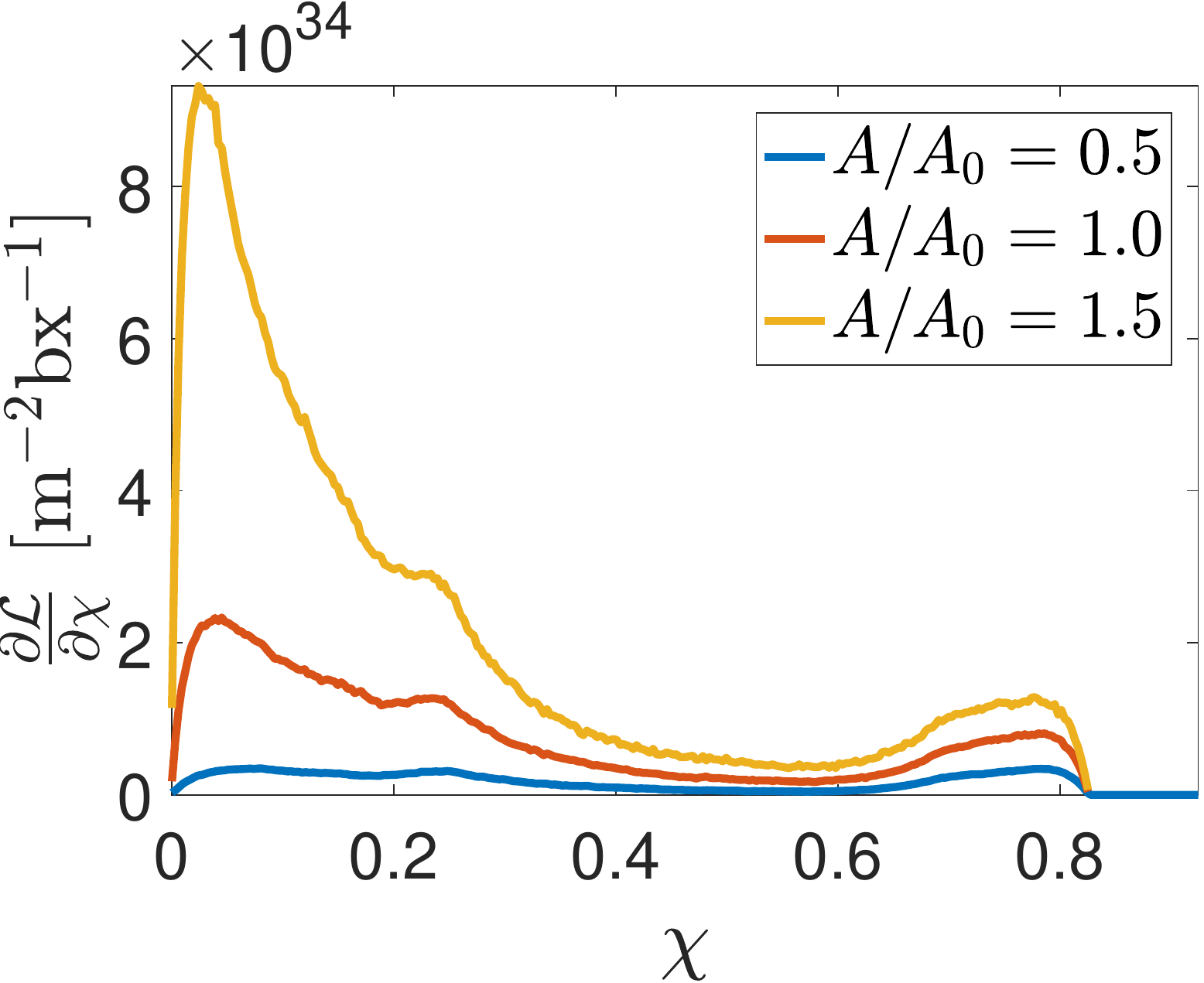}
    \caption{Luminosity spectra for $\gamma\gamma$ collisions for a laser with several target thicknesses $A/A_0$ and the distance $\varrho=1$ between the conversion region and interaction point. $\chi=\sqrt{s}/\sqrt{s_0}$.}
    \label{2020_N5e+09_sigmaZ5_betaY0147mm_betaX9mm_2hP-0.8_targThick_rho1_lumiSpec}
\end{figure}

\begin{figure}[!htb]
    \centering
    \includegraphics[width=0.65\columnwidth]{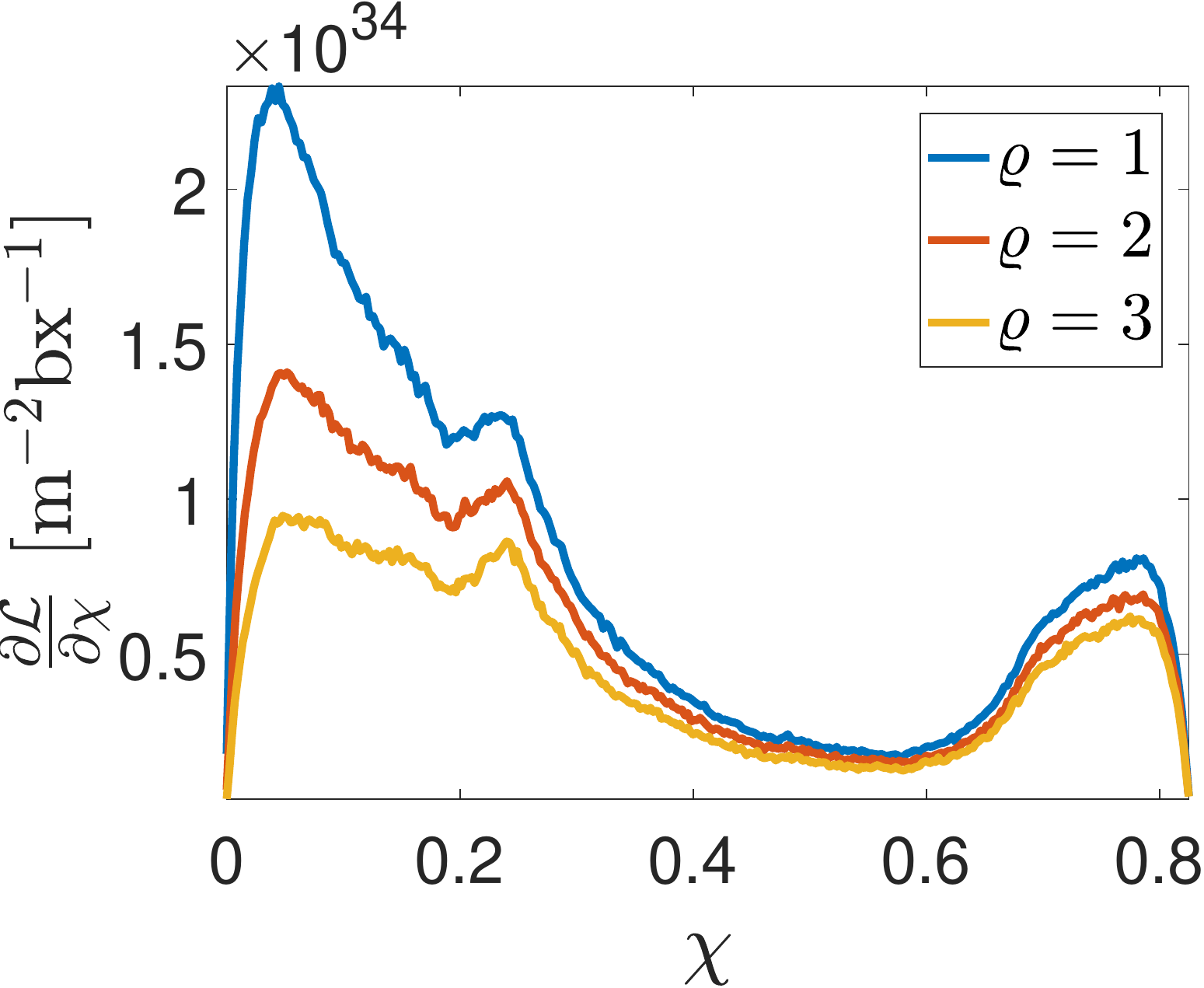}
    \caption{Luminosity spectra for $\gamma\gamma$ collisions for a laser with target thickness $A/A_0=1$ and several distances $\varrho$ between the conversion region and interaction point. $\chi=\sqrt{s}/\sqrt{s_0}$.}
    \label{2020_N5e+09_sigmaZ5_betaY0147mm_betaX9mm_2hP-0.8_targThick1_rho_lumiSpec}
\end{figure}

Thus, the total and peak luminosity can be maximised by minimising $\beta_x^*$ and $\beta_y^*$. The CLIC \SI{3}{\tera\electronvolt} parameter set uses $\beta_x^*=\SI{6.9}{\milli\meter}$ and $\beta_y^*=\SI{68}{\micro\meter}$, which represent the smallest beta functions currently achievable. These values for $\beta_{x,y}^*$ however take into account non-linear effects and are not matched to the spot sizes and emittances given in \cite{CLIC_CDR}, so we instead choose to assess the matched values $\beta_x^*=\SI{9.0}{\milli\meter}$ and $\beta_y^*=\SI{0.147}{\milli\meter}$ in this study. We performed additional GUINEA-PIG simulations for $\gamma\gamma$ collisions of photons scattered at electron beams with $N=\SI{5e9}{}$ electrons, $\sigma_z=\SI{5}{\micro\meter}$, $\beta_x^*=\SI{9.0}{\milli\meter}$, $\beta_y^*=\SI{0.147}{\milli\meter}$ to study the effect of $A/A_0$ and $\varrho$ on the luminosity spectrum. The luminosity spectra are plotted against $\chi=\sqrt{s}/\sqrt{s_0}$ in fig. \ref{2020_N5e+09_sigmaZ5_betaY0147mm_betaX9mm_2hP-0.8_targThick_rho1_lumiSpec} and \ref{2020_N5e+09_sigmaZ5_betaY0147mm_betaX9mm_2hP-0.8_targThick1_rho_lumiSpec}.

As expected, large laser pulse energies lead to an enhancement of the low-energy part of the luminosity spectrum due to increased chance of multiple scattering and an increased number of low-energy photons. For $A/A_0=0.5$, $\mathcal{L}_\mathrm{peak}/\mathcal{L}=0.25$ while $\mathcal{L}_\mathrm{peak}/\mathcal{L}=0.085$ for $A/A_0=1.5$. Furthermore, larger values of $\varrho$ result in smaller total luminosities, but enhance the relative sharpness of the spectrum with $\mathcal{L}_\mathrm{peak}/\mathcal{L}=0.14$ for $\varrho=1.0$ and $\mathcal{L}_\mathrm{peak}/\mathcal{L}=0.20$ for $\varrho=3.0$. Due to this dependence on $\varrho$, for a fixed distance $d$ between the conversion point and the IP, decreasing $\beta_y^*$ would lead to an improvement of the sharpness of the luminosity spectrum, but the increase in total luminosity would be less than the $1/\sqrt{\beta_y^*}$-scaling.

\subsection*{Background}
In this background study, we will consider the case where no beam separation scheme for the e\textsuperscript{$-$} beams are applied, and compare the results for two sets of $\beta_x^*$ and $\beta_y^*$. Set 1 contains the previously used values $\beta_x^*=\SI{9.0}{\milli\meter}$ and $\beta_y^*=\SI{0.147}{\milli\meter}$, while set 2 contains the smallest values $\beta_x^*=\SI{1.0}{\milli\meter}$ and $\beta_y^*=\SI{1.0}{\micro\meter}$ that were considered in this parameter scan. Other parameters such as $N=\SI{5e9}{}$ electrons, $\sigma_z=\SI{5}{\micro\meter}$, $2h_\mathrm{e}\Omega=-0.8$, $A/A_0=1.0$, and $\varrho=1.0$ are kept identical for parameter set 1 and 2. 

\begin{figure}[!htb]
    \centering
    \includegraphics[width=0.75\columnwidth]{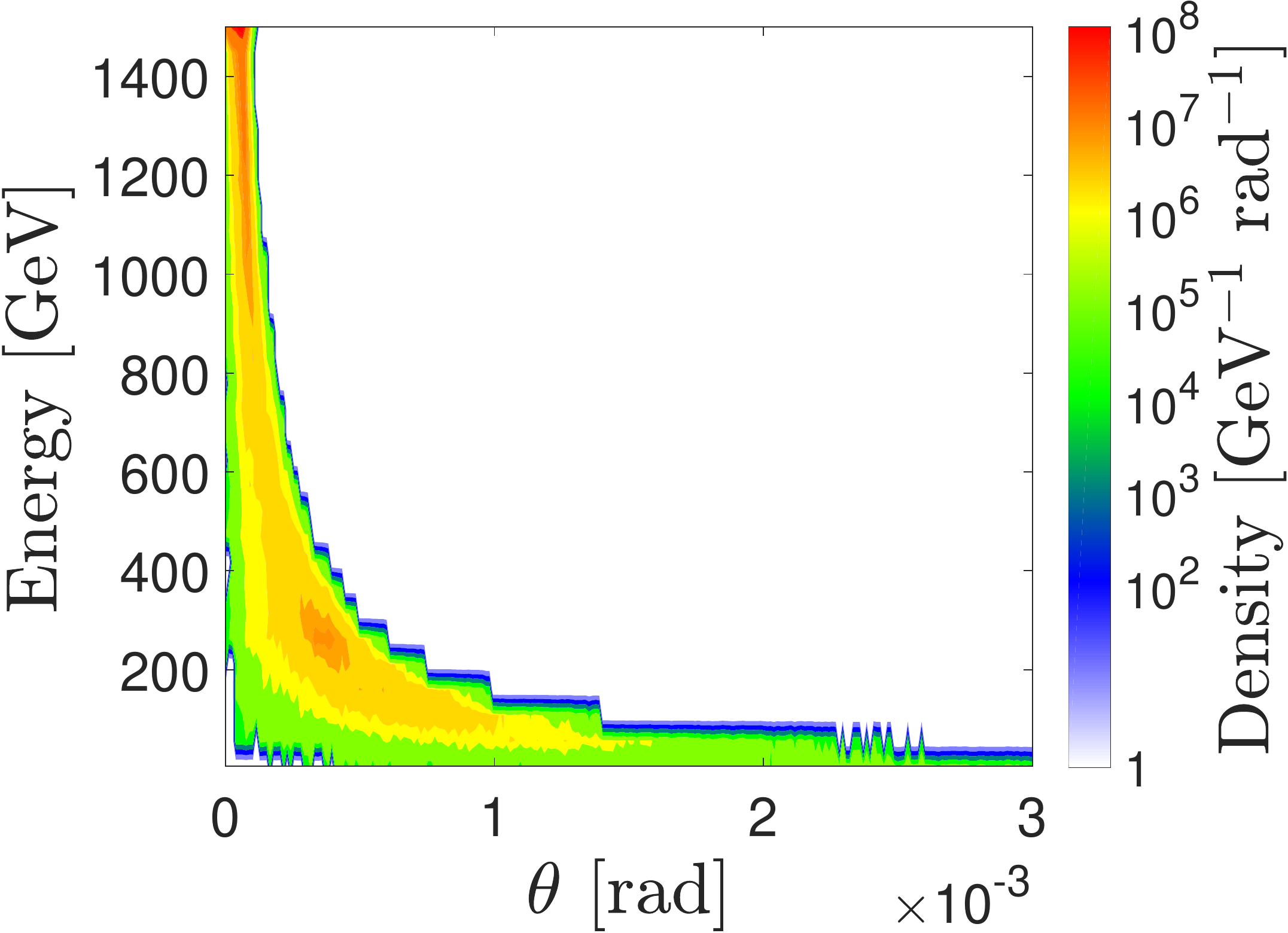}
    \caption{Energy-angular distribution of e\textsuperscript{$-$} in the spent beam for parameter set 1. $\theta$ is the angle with respect to the beam axis.}
    \label{fig:contourPlot_energy_thetar_2020_beam1_N5e+09_sigmaZ5_betaY68_betaX6-9mm}
\end{figure}
\begin{figure}[!htb]
    \centering
    \includegraphics[width=0.75\columnwidth]{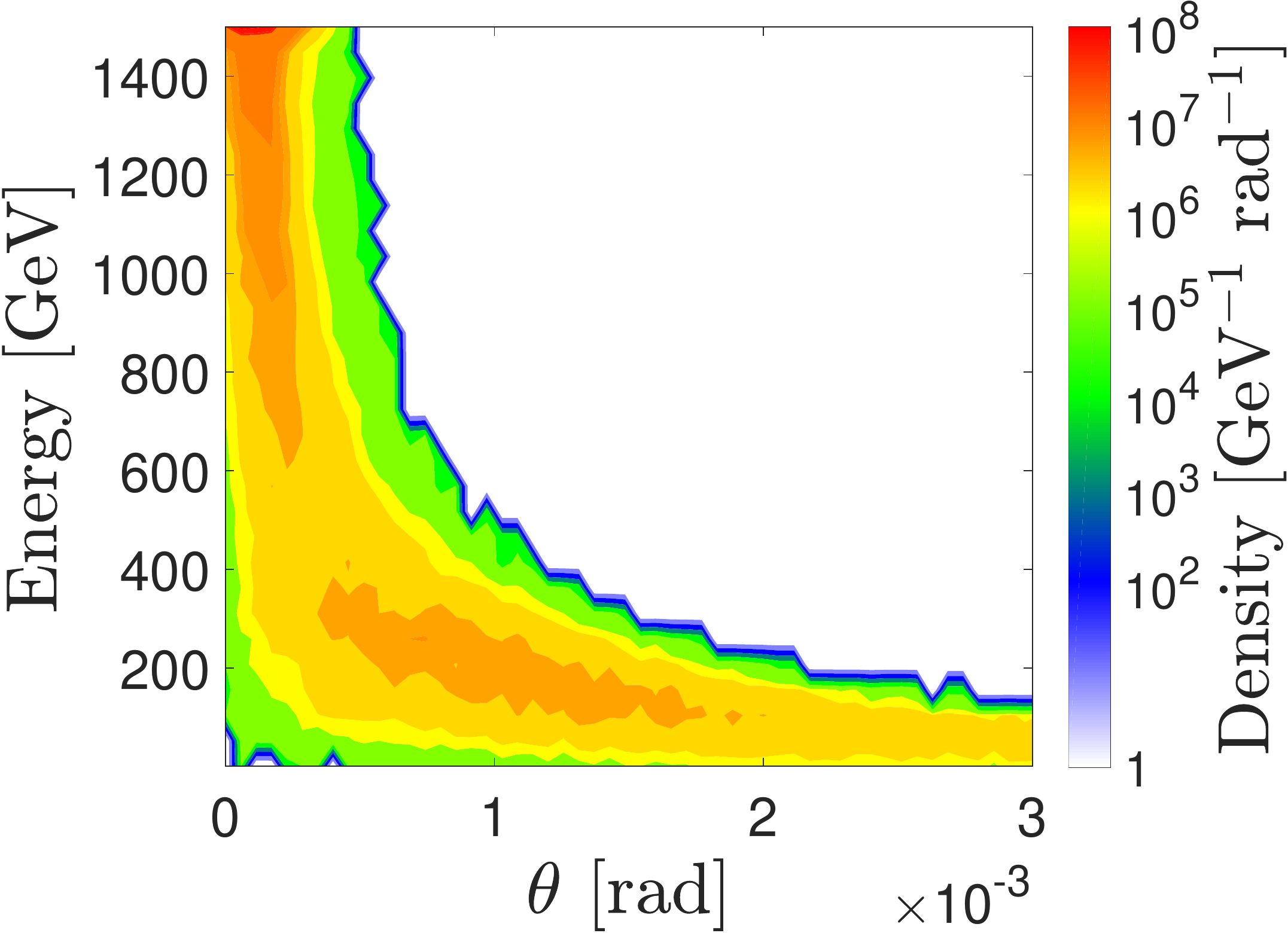}
    \caption{Energy-angular distribution of e\textsuperscript{$-$} in the spent beam for parameter set 2. $\theta$ is the angle with respect to the beam axis.}
    \label{fig:contourPlot_energy_thetar_2020_beam1_N5e+09_sigmaZ5_betaY1_betaX1mm}
\end{figure}

Contour plots of the energy-angular distribution for parameter set 1 is shown in fig. \ref{fig:contourPlot_energy_thetar_2020_beam1_N5e+09_sigmaZ5_betaY68_betaX6-9mm} and the same is shown for parameter set 2 in fig. \ref{fig:contourPlot_energy_thetar_2020_beam1_N5e+09_sigmaZ5_betaY1_betaX1mm}. $\theta$ in fig. \ref{fig:contourPlot_energy_thetar_2020_beam1_N5e+09_sigmaZ5_betaY68_betaX6-9mm} and \ref{fig:contourPlot_energy_thetar_2020_beam1_N5e+09_sigmaZ5_betaY1_betaX1mm} is the angle with respect to the beam axis.



The smaller beta functions in parameter set 2 clearly lead to a larger fraction of low energy particles propagating at large angles. This is likely caused by a higher number of scattering with the laser photons due to the higher beam density in addition to being deflected more severely by the opposite electron beam. The low energy e\textsuperscript{$-$} will be strongly deflected by the other incoming e\textsuperscript{$-$} beam, and thus the spent beam in parameter set 2 will lead to stronger background signals. 

In addition to deflected e\textsuperscript{$-$} from the spent beams, secondary particles can also contribute to the background signal. This include e\textsuperscript{$+$}e\textsuperscript{$-$} pair creation and hadrons produced by the interaction of the spent beams \cite{Schulte_thesis}. Table \ref{tab:comparison_lumi_background} summarises key figures such as the number $N_\mathrm{H}$ of hadronic events per beam crossing with a centre of mass energy above \SI{5}{\giga\electronvolt}, in addition to the number of coherent ($N_\mathrm{coh}$), trident ($N_\mathrm{tri}$) and incoherent ($N_\mathrm{inc}$) e\textsuperscript{$+$}e\textsuperscript{$-$} pairs produced per beam crossing. Various types of luminosities and energies are also listed. We also performed $\gamma\gamma$ collider simulations in GUINEA-PIG using the CLIC parameter set, and have included the results for comparison. As expected, parameter set 2 results in the highest $\gamma\gamma$ luminosity $\mathcal{L}_{\gamma\gamma}$ and also has the sharpest $\gamma\gamma$ luminosity spectrum.

$\mathcal{L}_{\gamma\gamma}$ of set 1 is larger than expected compared to the CLIC set, which should only have been a factor $\sim 1.56$ larger due to the ${N^2/\sqrt{\varepsilon_x}}$ scaling of luminosity. In order to examine if this is caused by the difference in ${\sigma_z}$, we performed a separate simulation using set 1 parameters except replacing ${\sigma_z=\SI{5}{\micro\meter}}$ with ${\sigma_z=\SI{44}{\micro\meter}}$. This separate simulation resulted in ${\mathcal{L}_{\gamma\gamma}=\SI{8.48e35}{\meter^{-2} bx^{-1}}}$ and ${\mathcal{L}_{\gamma\gamma}^\mathrm{peak}=\SI{1.27e35}{\meter^{-2} bx^{-1}}}$, which agree well with the ${N^2/\sqrt{\varepsilon_x}}$-scaling and indicate a $\sigma_z$ dependence for large $\sigma_z$.

Compared to $\mathcal{L}_{\gamma\gamma}$, both the $\gamma\mathrm{e}^-$ luminosity $\mathcal{L}_{\gamma\mathrm{e}^-}$ and the e\textsuperscript{$-$}e\textsuperscript{$-$} luminosity $\mathcal{L}_{\mathrm{e}^-\mathrm{e}^-}$ are 2--3 orders of magnitude lower for all three parameter sets. The number of background particles for parameter set 2 is higher than the other sets, which is expected due to more intense beam fields leading to the production of more e\textsuperscript{$-$}e\textsuperscript{$-$} pairs. The total energies of the background particles are also highest for parameter set 2. Furthermore, $N_\mathrm{H}$ is also significantly higher for parameter set 2, so that it overall produces the strongest background. 

\begin{table}[!hbt]
\centering
\caption{Luminosity and Background Comparison}
    \begin{tabular}{llccc}
    \toprule
        \textbf{Quantity}  &   \textbf{Unit}  &   \textbf{Set 1}  &   \textbf{Set 2}  &   \textbf{CLIC}\\
    \midrule
    $N$ &   $10^9$  &   5.0 &   5.0 &   3.72\\
    
    $\sigma_z$  &   \SI{}{\micro\meter} &   5.0 &   5.0 &   44.0\\
    
    $\beta_x^*/\beta_y^*$   &   mm/\SI{}{\micro\meter}  &   9.0/147 &    1.0/1.0 &   9.0/147\\
    
    $\gamma\varepsilon_x$   &   \SI{}{\milli\meter \milli\radian}  &   0.887  &   0.887  &0.66\\
    
    $\gamma\varepsilon_y$   &   \SI{}{\milli\meter \milli\radian}  &   0.02  &   0.02  &   0.02\\
    
    $\mathcal{L}_{\gamma\gamma}$   &   \SI{e35}{\meter^{-2}} $\mathrm{bx}^{-1}$ &  12.09 &  209.16 &   5.45\\
    
    $\mathcal{L}_{\gamma\gamma}^\mathrm{peak}$   &   \SI{e35}{\meter^{-2}} $\mathrm{bx}^{-1}$ &  1.72 &  41.51 &   0.85\\
    
    $\mathcal{L}_{\gamma\mathrm{e}^-}$   &   \SI{e35}{\meter^{-2}} $\mathrm{bx}^{-1}$ &   0.06 &   1.08 &   0.03\\
    
    $\mathcal{L}_{\mathrm{e}^-\mathrm{e}^-}$   &   \SI{e35}{\meter^{-2}}$\mathrm{bx}^{-1}$ &   0.05    &   0.56    &   0.02\\
    
    $N_\mathrm{coh}$    &   $10^8$  &   1.73   &    13.49   & 0.08\\
    
    $\calE_\mathrm{coh}$   &   \SI{e11}{\giga\electronvolt}   &   0.33  &   1.66  &    0.03\\
    
    $N_\mathrm{tri}$    &   $10^7$  &   0.91 &  5.45   &   0.04\\
    
    $\calE_\mathrm{tri}$    &   \SI{e9}{\giga\electronvolt} &   1.03    &   2.70 &  0.10\\
    
    $N_\mathrm{inc}$     &   $10^5$  &   0.26   &   3.46   &   0.12\\
    
    $\calE_\mathrm{inc}$   &   \SI{e7}{\giga\electronvolt}   &   0.12  &   2.12  &    0.06\\
    
    $N_\mathrm{H}$  &   &   70.15    &   1291.33    &   33.17\\
    
    \bottomrule
    \end{tabular}
\label{tab:comparison_lumi_background}
\end{table}


\section{CONCLUSION}
The absence of beam-beam effects in $\gamma\gamma$ collisions allows for increasing the total luminosity by minimising $\beta_{x,y}^*$ of incoming electron beams used for back-scattering photons. Electron beams with smaller $\beta_{x,y}^*$ however also lead to stronger backgrounds that can disrupt the experimental results. 

The combined parameter set consisting of PWFA beam parameters and smallest currently achievable $\beta_{x,y}^*$ was able to provide a significantly higher total luminosity than a $\gamma\gamma$ collider based on CLIC parameters, but also produces a larger background. A detailed study of the detector design is required in order to ensure that the presented backgrounds are within acceptable levels.



%
%
\ifboolexpr{bool{jacowbiblatex}}%
	{\printbibliography}%

\begin{thebibliography}{9} 
	
    \bibitem{Blumenfeld}
	    I. Blumenfeld \textit{et al.},
	    `Energy doubling of 42 GeV electrons in a metre-scale plasma wakefield accelerator',
	    \textit{Nature},
	    vol. 445,
	    pp. 741--4,
	    Mar. 2007.
	    
    \bibitem{Telnov_gg_Physics_Goal}
        V. I. Telnov,
        `Physics goals and parameters of photon colliders',
        \textit{Int. J. Mod. Phys. A},
        vol. 13,
        pp. 2399--2410,
        1988.
	    
    \bibitem{Chen2020modeling}
        J. B. B. Chen, D. Schulte and E. Adli, 
        `Modeling and simulation of transverse wakefields in PWFA',
        \textit{J. Phys. Conf. Ser.},
        vol. 1596,
        p. 012057,
        Jul. 2020.
        
    \bibitem{Telnov_gammagammaPrinciples}
        V. I. Telnov,
        `Principles of photon colliders',
        \textit{Nucl. Instr. Meth. Phys. Res. Sec. A},
        vol. 355,
        no. 1,
        pp. 3-18,
        1995
        
    \bibitem{Ginzburg_gammagamma}
        I. F. Ginzburg, G. L. Kotkin, V. G. Serbo and V. I. Telnov,
        `Colliding ge and gg beams based on the single-pass e\textsuperscript{$\pm$}e\textsuperscript{--} colliders ({VLEPP} type)',
        \textit{Nucl. Instr. Meth. Phys. Res.,}
        vol. 205,
        no. 1,
        pp. 47-68,
        1983.
    
    \bibitem{Schulte_thesis}
        D. Schulte, 
        `Study of Electromagnetic and Hadronic Background in the Interaction Region of the TESLA Collider',
        PhD thesis,
        Hamburg U.,
        1997.
    
    \bibitem{Chen2020ee}
    J. B. B. Chen, D. Schulte and E. Adli,
    \textit{{e\textsuperscript{$+$}e\textsuperscript{$-$}} beam-beam parameter study for a {TeV}-scale {PWFA} linear collider},
    2020.
    arXiv:\texttt{2009.13672}
    
    \bibitem{CLIC_CDR}
        M. Aicheler \textit{et al.},
        \textit{A Multi-TeV Linear Collider Based on CLIC Technology: CLIC Conceptual Design Report},
        ser. CERN Yellow Reports: Monographs.
        Geneva: CERN,
        2012.
        
        
	\end{thebibliography}
	{%
	

} 
%
%


\end{document}